\documentclass[twocolumn,prd,superscriptaddress,showpacs,floatfix,%
preprintnumbers,nofootinbib]{revtex4}

\usepackage{epsfig}

\newcommand{\D}{{\rm d}}

\begin{document}

\title{Self-induced parametric resonance in collective neutrino
oscillations}

\author{Georg G.~Raffelt}
\affiliation{Max-Planck-Institut f\"ur Physik
(Werner-Heisenberg-Institut), F\"ohringer Ring 6, 80805 M\"unchen,
Germany}

\date{8 October 2008}

\preprint{MPP-2008-113}

\begin{abstract}
We identify a generic new form of collective flavor oscillations in
dense neutrino gases that amounts to a self-induced parametric
resonance. It occurs in a homogeneous and isotropic ensemble when a
range of neutrino modes is prepared in a different flavor than the
neighboring modes with lower and higher energies. The flavor content
of the intermediate spectral part librates relative to the other
parts with a frequency corresponding to a typical $\Delta m^2/2E$.
This libration persists in the limit of an arbitrarily large
neutrino density where one would have expected synchronized flavor
oscillations.
\end{abstract}


\maketitle

\section{Introduction}                        \label{sec:introduction}

Flavor oscillations in dense neutrino gases exhibit collective
phenomena caused by neutrino-neutrino
interactions~\cite{Pantaleone:1992eq, Pantaleone:1998xi, Sigl:1992fn,
Samuel:1993uw, Kostelecky:1993dm, Kostelecky:1995dt, Samuel:1996ri,
Pastor:2001iu, Dolgov:2002ab, Wong:2002fa, Abazajian:2002qx,
Pastor:2008ti, Pastor:2002we, Sawyer:2004ai, Sawyer:2005jk,
Sawyer:2008zs, Bell:2003mg, Friedland:2003eh, Friedland:2003dv,
Friedland:2006ke, Balantekin:2006tg, Duan:2005cp, Duan:2006an,
Hannestad:2006nj, Raffelt:2007yz, EstebanPretel:2007ec, Duan:2007mv,
Raffelt:2007cb, Raffelt:2007xt, Duan:2007fw, Duan:2007bt,
Fogli:2007bk, Fogli:2008pt, Duan:2007sh, Dasgupta:2008cd,
EstebanPretel:2007yq, Dasgupta:2007ws, Duan:2008za, Dasgupta:2008my,
EstebanPretel:2008ni, Chakraborty:2008zp, Gava:2008rp}.  In the
simplest example of a dense neutrino gas that is homogeneous and
isotropic, studies of collective oscillations in the two-flavor
context amount to solving the nonlinear equations of motion (EOMs)
for a set of flavor polarization vectors~${\bf P}_i$,
\begin{equation}\label{eq:EOM1}
\dot {\bf P}_i={\bf H}_i\times {\bf P}_i\,,
\end{equation}
where
\begin{equation}\label{eq:ham1}
{\bf H}_i=\omega_i{\bf B}+\mu{\bf P}\,.
\end{equation}
The ``effective magnetic field'' ${\bf B}$ is a unit vector in flavor
space and $\omega_i=\Delta m^2/2E_i$ the vacuum oscillation frequency
for a neutrino mode with energy $E_i$. The total polarization vector
of the ensemble~is
\begin{equation}
{\bf P}=\sum_i {\bf P}_i\,.
\end{equation}
Finally, $\mu\sim\sqrt2\,G_{\rm F}n_\nu$ is a typical
neutrino-neutrino interaction energy. Its exact definition depends
on how we normalize the polarization vectors, but for our purpose it
is simply an adjustable parameter of dimension energy.

Two main questions are of interest: What are the collective forms of
motion in a dense neutrino gas? What happens when $\mu$ slowly
decreases, mimicking the expanding universe or the neutrino density
variation with distance from an astrophysical source? We concentrate
on the former question and identify a new mode of collective
oscillations that so far has gone unnoticed.

The simplest form of collective motion consists of ``synchronized
oscillations,'' meaning that modes with different frequencies
$\omega_i$ oscillate with a common frequency $\omega_{\rm sync}$ if
$\mu$ is sufficiently large~\cite{Samuel:1993uw, Kostelecky:1995dt,
Pastor:2001iu}. The main physical idea is that the Hamiltonian
vectors ${\bf H}_i$ in Eq.~(\ref{eq:ham1}) are dominated by $\mu
{\bf P}$, forcing all ${\bf P}_i$ to follow ${\bf P}$ and thus to a
common precession around~${\bf B}$. It is this picture that we will
see is not always complete.

Another case is the ``pure precession mode'' where each ${\bf P}_i$
is collinear (parallel or antiparallel) to ${\bf H}_i$
\cite{Duan:2007mv, Raffelt:2007cb, Raffelt:2007xt, Duan:2007fw}. Even
though ${\bf H}_i$ depends on all ${\bf P}_i$ through ${\bf P}$, such
a self-consistent solution exists for any strength of $\mu$. All
${\bf H}_i$ lie in the plane spanned by ${\bf B}$ and ${\bf P}$. This
plane rotates around ${\bf B}$, all ${\bf P}_i$ being static within
it. For $\mu\to\infty$ this solution requires all ${\bf P}_i$ to be
collinear and then is identical with a synchronized solution.
Conversely, beginning with collinear ${\bf P}_i$ and slowly reducing
$\mu$ takes us adiabatically through different pure precession modes.

Finally, the ``pendulum in flavor space'' represents a class of
solutions relevant for $\mu$ not too large~\cite{Duan:2005cp,
Hannestad:2006nj, Duan:2007mv}. The simplest case involves two
vectors ${\bf P}_{1,2}$ that initially point in opposite directions.
The dynamics of this system is equivalent to a gyroscopic pendulum.
Once more, for $\mu\to\infty$ the motion is a common synchronized
precession without nutation unless initially ${\bf P}={\bf P}_1+{\bf
P}_2=0$.

It was always assumed that, unless initially ${\bf P}=0$, the
large-$\mu$ behavior is a synchronized precession. Likewise, it was
assumed that the pure precession mode is stable. However, while a
small perturbation may lead to small oscillations around the ideal
solution, it may also lead to an exponential deviation. Numerical
studies always found stable behavior for the pure precession mode and
synchronized behavior in the large-$\mu$ limit. However, we will see
that these findings depend on special choices of initial conditions
and are not generic, although probably most relevant in the supernova
context.

It is easy to see that the pure precession mode and its large-$\mu$
limit need not be stable. A pure precession mode with each ${\bf
P}_i$ collinear with ${\bf H}_i$ is a self-consistent exact solution.
Realistic initial conditions, however, begin with all ${\bf P}_i$
collinear with each other, representing the assumption of all
neutrinos being prepared in weak-interaction eigenstates. On the
other hand, the initial $\mu$ may be large but must be finite. This
initial condition is not an exact pure precession mode because each
${\bf P}_i$ sports a small angle relative to its ${\bf H}_i$ and thus
moves on a precession cone with a small but nonzero opening angle.
Therefore the total ${\bf P}$ itself can not be static in the
co-rotating plane.  Each ${\bf P}_i$ precesses around ${\bf H}_i$
with an approximate frequency $\mu$, so ${\bf P}$ itself must
``vibrate'' with a similar frequency.

We thus have a typical situation for a parametric resonance: Each
${\bf P}_i$ precesses around ${\bf H}_i$ with the approximate
frequency $\mu$ and ${\bf H}_i$ itself vibrates with a similar
frequency. Therefore, while the precession cone of each ${\bf P}_i$
follows its ${\bf H}_i$, it is not assured that the opening angle
stays small. The exact outcome depends on how the evolution of the
${\bf P}_i$ feeds back on ${\bf P}$ and thus on ${\bf H}_i$.

To identify the simplest system showing such behavior
we return to the EOMs of Eq.~(\ref{eq:EOM1}) and note that they have
two exact invariants. One is the angular momentum along ${\bf B}$, so
${\bf B}\cdot{\bf P}$ is conserved~\cite{Hannestad:2006nj}. The other
is the energy~\cite{Duan:2005cp, Raffelt:2007yz}
\begin{equation}\label{eq:energy}
{\bf B}\cdot{\bf M}+\frac{\mu}{2}\,{\bf P}^2\,,
\end{equation}
where the ``total magnetic moment'' is
\begin{equation}
{\bf M}=\sum_i \omega_i{\bf P}_i\,.
\end{equation}
When $\mu\to\infty$ energy conservation implies that ${\bf P}^2$ and
thus $|{\bf P}|$ are conserved~\cite{Duan:2005cp}. ${\bf P}$ then
precesses around ${\bf B}$ as a collective object, representing the
usual synchronized oscillations.

This does not imply, however, that all ${\bf P}_i$ remain collinear
with ${\bf P}$. There can be internal motions among them such that
$|{\bf P}|$ is conserved and ${\bf P}$ still precesses as a
collective object, yet the individual ${\bf P}_i$ move relative to
each other and relative to ${\bf P}$.

For two polarization vectors no internal motion is possible without
modifying $|{\bf P}|$. Moreover, since $|{\bf P}_i|$ is conserved,
the motion of each ${\bf P}_i$ is described by two angles, so we have
a total of four degrees of freedom. With two exact constants of the
motion there remain two degrees of freedom, representing, for
example, the ``nutation angle'' and the ``precession angle'' of the
flavor pendulum. There is no room for new forms of motion.

If we have three or more polarization vectors that are initially
aligned, nothing new happens either because the approximate
conservation of $|{\bf P}|$ in the large-$\mu$ limit requires that
the ${\bf P}_i$ remain aligned.

Thus we need three polarization vectors (oscillation frequencies
$\omega_1<\omega_2<\omega_3$), one of them initially anti-aligned
with the others.  If the ``flipped'' polarization vector is number 1
or 3 we are back to the flavor pendulum because in the large-$\mu$
limit two neighboring aligned vectors act roughly as one average
mode. This is not possible if the flipped vector is number 2, so the
relevant initial configuration is either up-down-up or down-up-down.

The main purpose of this paper is to study the most basic system of
this sort and some simple generalizations. In Sec.~\ref{sec:three} we
solve analytically a very symmetric system consisting of three modes.
In Sec.~\ref{sec:spectrum} we consider its counterpart for a
continuous spectrum of modes and solve it analytically. We discuss
the implications of our findings in Sec.~\ref{sec:conclusions}.

\section{Three Polarization Vectors}                 \label{sec:three}

\subsection{Basic configuration}

The basic configuration showing a self-induced parametric resonance
consists of three polarization vectors with frequencies
$\omega_1<\omega_2<\omega_3$. For simplicity we assume equal
frequency spacings $\gamma=\omega_3-\omega_2=\omega_2-\omega_1$. The
common precession of the system around ${\bf B}$ is irrelevant, so we
can choose $\omega_2=0$, essentially going to a rotating frame around
${\bf B}$. The polarization vectors are ${\bf P}_{\pm,0}$ with
oscillation frequencies $\pm\gamma$ and $0$ and the EOMs are
\begin{eqnarray}
\dot{\bf P}_\pm&=&
(\pm\gamma{\bf B}+\mu{\bf P})\times{\bf P}_\pm\,,
\nonumber\\
\dot{\bf P}_0\;&=&\mu{\bf P}\times{\bf P}_0\,.
\end{eqnarray}
Moreover, we assume that all polarization vectors have unit length and
that initially ${\bf P}_\pm$ point in the positive,
${\bf P}_0$ in the negative $z$-direction.

\subsection{Numerical examples}

\begin{figure}[b]
\includegraphics[width=0.8\columnwidth]{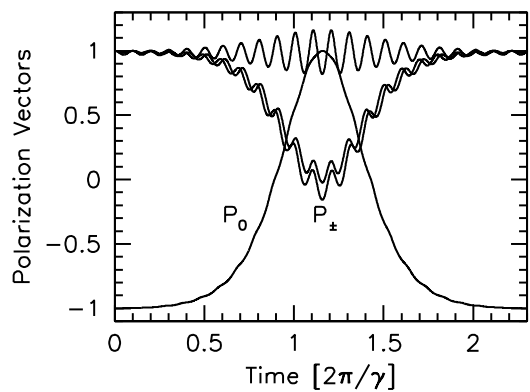}
\caption{Evolution of three polarization vectors for
$\mu=10\,\gamma$.
We show ${\bf P}_\pm$ and ${\bf P}_0$, projected on the direction of
the total ${\bf P}$. The uppermost curve is $|{\bf P}|$.
\label{fig:threeA}}
\end{figure}

We first illustrate the behavior of this system with a numerical
example. We take ${\bf B}$ to be tilted relative to the $z$-axis by
an angle $2\theta$ with $\cos2\theta=0.5$, $\theta$ itself playing
the role of the neutrino mixing angle. In Fig.~\ref{fig:threeA} we
show the evolution of the projections of ${\bf P}_\pm$ and ${\bf
  P}_0$ on ${\bf P}$ for the interaction strength
$\mu=10\,\gamma$. The vector ${\bf P}_0$ evolves from its initial
anti-alignment with ${\bf P}$ to complete alignment and back, and so
forth periodically. The other two vectors evolve similar to each
other such that ${\bf B}\cdot{\bf P}$ is strictly conserved and the
length $|{\bf P}|$ (uppermost curve) is approximately conserved.

\begin{figure}
\includegraphics[width=0.8\columnwidth]{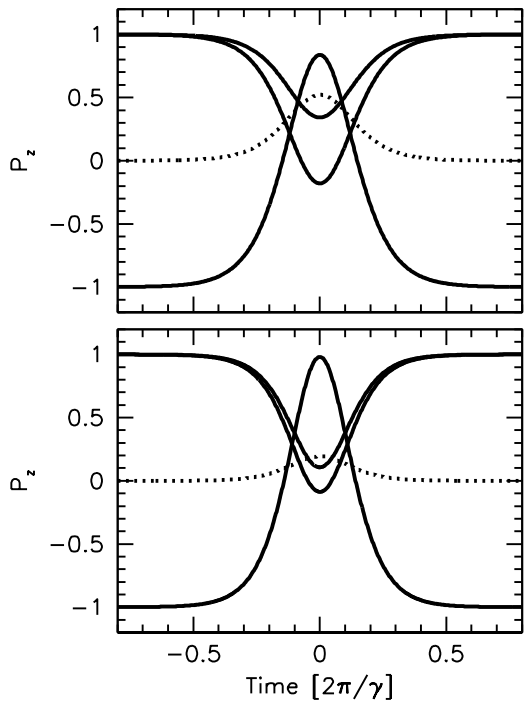}
\caption{Evolution of three polarization vectors as in
Fig.~\ref{fig:threeA}, now with a small mixing angle and
$\mu=3\,\gamma$ (upper panel) and $\mu=10\,\gamma$ (lower panel). The
dotted line corresponds to the interaction energy
(see text).\label{fig:threeB}}
\end{figure}

The interesting internal evolution is relative to ${\bf P}$, so a
large $\theta$ complicates the system by projection effects. We may
use the opposite extreme of a vanishing $\theta$ because the
existence of the new collective mode can not depend on this choice. A
vanishing $\theta$ has the only disadvantage that the system can not
start moving if the polarization vectors begin exactly collinear with
${\bf B}$. However, the motion can be excited by using a small but
nonzero $\theta$ or by some other disturbance.

In Fig.~\ref{fig:threeB} we show the evolution for an extremely small
but nonzero mixing angle using $\mu=3\gamma$ (upper panel) and
$\mu=10\gamma$ (lower panel). We do not show the long quasi-static
phase where the system starts moving. One difference to the previous
case is the absence of any ``wiggles'' even for a moderate $\mu$. For
larger $\mu$ the variation of the interaction energy $(\mu/2){\bf
P}^2$ becomes smaller. We show as a dotted line the quantity
$(\mu/2)\,({\bf P}^2-1)$. For $\mu\to\infty$ the two energy
components ${\bf B}\cdot{\bf M}$ and $(\mu/2)\,{\bf P}^2$ seem to be
separately conserved. It is striking that the ``flipping time scale''
remains the same.

\begin{figure}
\includegraphics[width=1.0\columnwidth]{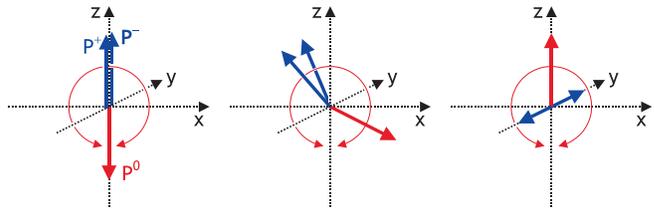}
\caption{Internal motion of three
  polarization vectors.
\label{fig:libration}}
\end{figure}

We sketch the motion of the polarization vectors in
Fig.~\ref{fig:libration}.  The vector ${\bf P}_0$ oscillates relative
to ${\bf P}$, its relative orientation being described by the angle
$\varphi$ relative to the negative ${\bf P}$ direction. If we use
coordinates where the $x$-direction is defined by ${\bf P}_0$ we can
describe the motion by the angle $\varphi$ alone in the form
\begin{equation}\label{eq:sol1}
{\bf P}_0=\pmatrix{s\cr 0\cr -c\cr}
\quad{\rm and}\quad
{\bf P}_\pm=\frac{1}{2}
\pmatrix{-s\cr \pm\sqrt{2(1-c)}\cr1+c\cr}\,,
\end{equation}
where $s=\sin\varphi$ and $c=\cos\varphi$. In this way we always have
${\bf P}=(0,0,1)$. The energy components ${\bf B}\cdot{\bf M}$ and
$(\mu/2){\bf P}^2$ do not depend on~$\varphi$.

\subsection{Analytic solution}

Equation~(\ref{eq:sol1}) must be a good approximation to the
solutions of the EOMs, so all that is missing is the time evolution
of $\varphi$. To find it we simplify the EOMs further and observe
that the component of ${\bf P}$ along ${\bf B}$ is strictly
conserved. Our choice of unit length for all vectors therefore
implies ${\bf P}={\bf B}+{\bf
  P}_\perp$ where ${\bf P}_\perp$ is the ${\bf P}$ component
transverse to ${\bf B}$. Therefore, the EOMs are
\begin{eqnarray}
\dot{\bf P}_\pm&=&\mu{\bf P}_\perp\times{\bf P}_\pm+
   (\mu\pm\gamma)\,{\bf B}\times{\bf P}_\pm\,,
\nonumber\\
\dot{\bf P}_0\;&=&\mu{\bf P}_\perp\times{\bf P}_0\;+
\mu{\bf B}\times{\bf P}_0\,.
\end{eqnarray}
We observe that there is a common precession with frequency $\mu$
around ${\bf B}$ that represents the fast internal precession of all
${\bf P}_i$ around ${\bf P}$. We can take out this common motion by
going once more to a rotating frame so that finally the EOMs for the
internal motion are
\begin{eqnarray}\label{eq:EOM5}
\dot{\bf P}_\pm&=&\mu{\bf P}_\perp\times{\bf P}_\pm
   \pm\gamma{\bf B}\times{\bf P}_\pm\,,
\nonumber\\
\dot{\bf P}_0\;&=&\mu{\bf P}_\perp\times{\bf P}_0\,.
\end{eqnarray}
We have achieved that fast precessions that are common to all
polarization vectors have disappeared. The polarization vectors see
the usual ${\bf B}$, leading to normal precessions, and in addition a
transverse effective field represented by ${\bf P}_\perp$ that tilts
them.

This is analogous to the textbook example of magnetic resonance where
an electron spin precesses fast in a strong external $B$ field.
Applying even the tiniest transverse $B$ field that rotates with the
original precession frequency leads to a situation where in the
co-rotating frame the spin only feels the transverse component and
completely reverses relative to the direction of the large external
field.

In this example the rotation speed of the transverse $B$ component
must be adjusted to match the electron's precession frequency. In our
case this resonance condition arises automatically by the nonlinear
nature of the problem where the transverse field is provided by the
precessing polarization vectors themselves.

To solve the EOMs it is useful to introduce the difference vector
\begin{equation}
{\bf D}={\bf P}_+-{\bf P}_-
\end{equation}
and study the motion of the three vectors ${\bf P}_0$, ${\bf D}$ and
${\bf P}_\perp$. Their EOMs are
\begin{eqnarray}
\label{eq:EOM3a}
\dot{\bf P}_0\kern0.2em&=&\mu{\bf P}_\perp\times{\bf P}_0\,,
\\*
\label{eq:EOM3b}
\dot{\bf D}\kern0.6em&=&\mu{\bf P}_\perp\times{\bf D}
+\gamma{\bf B}\times({\bf P}_\perp-{\bf P}_0)\,,
\\*
\label{eq:EOM3c}
\dot{\bf P}_\perp&=&\gamma{\bf B}\times
\left({\bf D}-\frac{\mu}{\gamma}\,{\bf P}_\perp\right)\,.
\end{eqnarray}
The vectors ${\bf P}_0$ and ${\bf D}$ have lengths of order unity,
whereas in the large-$\mu$ limit ${\bf P}_\perp$ must be small.
Energy conservation implies that it must be of order
$(\gamma/\mu)^{1/2}$ or smaller. However, since ${\bf P}_\perp$
appears on both the l.h.s.\ and r.h.s.\ of Eq.~(\ref{eq:EOM3c}) and
since it appears multiplied with $\mu$ on the r.h.s., both sides of
this equation are of the same order in $\gamma/\mu$ only if to
leading order
\begin{equation}
{\bf P}_\perp=\frac{\gamma}{\mu}\,{\bf D}\,.
\end{equation}
Here we have assumed that to leading order ${\bf D}$ has no component
along ${\bf B}$ because ${\bf P}_+$ and ${\bf P}_-$ are expected to
behave symmetrically. Inserting this into Eqs.~(\ref{eq:EOM3a})
and~(\ref{eq:EOM3b}) and neglecting terms of order $\gamma/\mu$ we
find
\begin{eqnarray}
\label{eq:EOM4a}
\dot{\bf P}_0&=&\kern0.7em\gamma{\bf D}\times{\bf P}_0\,,
\nonumber\\*
\label{eq:EOM4b}
\dot{\bf D}\kern0.4em&=&-\gamma{\bf B}\times{\bf P}_0\,.
\end{eqnarray}
To leading order in $\gamma/\mu$ we have found a closed system of
EOMs for ${\bf P}_0$ and ${\bf D}$ alone. The strength of the
neutrino-neutrino interaction has disappeared, only $\gamma$ appears
as an energy scale.

In our coordinate system ${\bf P}_0$ moves in the $x$-$z$ plane
whereas ${\bf D}$ is along the $y$-direction. Describing the
orientation of ${\bf P}_0$ with the angle $\varphi$ as in
Eq.~(\ref{eq:sol1}), the EOMs finally become
\begin{eqnarray}
\label{eq:EOM5a}
\dot\varphi&=&\gamma D\,,
\nonumber\\*
\label{eq:EOM5b}
\dot D&=&-\gamma\sin\varphi\,.
\end{eqnarray}
We recognize that $\varphi$ and $D$ play the role of canonically
conjugate variables and that these EOMs follow from a Hamiltonian
\begin{equation}
H(\varphi,D)=\gamma
\left[{\textstyle\frac{1}{2}}\,D^2+(\cos\varphi-1)\right]
\end{equation}
by virtue of $\dot\varphi=\partial H/\partial D$ and $\dot
D=-\partial H/\partial\varphi$. Therefore, we have the EOM of a
pendulum with frequency~$\gamma$.

The initial condition ${\bf P}_\pm$ aligned with ${\bf B}$ and ${\bf
P}_0$ anti-aligned corresponds to the initial condition $D=0$ and
$\varphi=0$. In other words, the pendulum begins in an inverted
position and obeys
\begin{equation}
\frac{\dot\varphi^2}{2}=\gamma^2\,(1-\cos\varphi)\,.
\end{equation}
This agrees well with numerical examples.

Using the vectors of Eq.~(\ref{eq:sol1}) with any value $\varphi_0$
as an initial condition amounts to $D_0=[2(1-\cos\varphi_0)]^{1/2}$
or $\dot\varphi_0=\gamma\,[2(1-\cos\varphi_0)]^{1/2}$. The system
follows the same solution (up to corrections of order $\gamma/\mu$)
for any $\varphi_0$, except that the pendular motion begins in some
phase of the oscillation other than the inverted position. The
pendulum begins with an initial excursion $\varphi_0$ and an initial
velocity $\dot\varphi_0$, both adjusted such that it will reach the
inverted position.

We can also choose other initial conditions with an arbitrary
$\varphi_0$ and no initial velocity ($D_0=0$). In this case we have a
pendulum that oscillates between two maximal excursion angles without
reaching the inverted position. In terms of the polarization vectors
this corresponds to ${\bf P}_\pm$ being initially aligned with ${\bf
B}$ whereas ${\bf P}_0$ begins with a nonvanishing angle $\varphi_0$.

We can choose this initial angle such that all three polarization
vectors are initially almost aligned. In this case we obtain
small-amplitude oscillations. This shows that our solution is also
relevant for three initially aligned polarization vectors, the case
where one finds synchronized oscillations. In the limit
$\mu\to\infty$ and exactly aligned vectors nothing happens. However,
for a finite $\mu$ and a small disturbance these vectors do
oscillate. Here the small opening angles of the precession cones
caused by the imperfect initial alignment between ${\bf P}_i$ and
${\bf H}_i$ also cause a parametric resonance, but in the direction
of the small opening angles becoming smaller, then back to the
original angle, and so forth. So the parametric resonance always
occurs, but has a visible effect only if ${\bf P}_0$ is initially
anti-aligned with ${\bf P}_\pm$.

We finally remark that the initial orientation of the three
polarization vectors relative to ${\bf B}$ is irrelevant. The initial
conditions up-down-up or down-up-down are equivalent. In contrast to
the gyroscopic flavor pendulum the neutrino mass hierarchy is here
irrelevant.

\subsection{Large mixing angle}

We return to the case where ${\bf P}_\pm$ and ${\bf P}_0$ are
initially at some large angle $2\theta$ relative to ${\bf B}$,
corresponding to a large neutrino mixing angle $\theta$. Armed with
the insights gained in the previous section we note that we should
study this situation in a coordinate system co-moving with ${\bf P}$
and co-rotating with ${\bf P}_0$ around ${\bf P}$. Now the vector
${\bf B}$ rotates with the large frequency $\mu$ around ${\bf P}$,
its transverse component averaging to zero. In other words, we may
consider the rotation-averaged EOMs in the same spirit as for the
ordinary matter effect~\cite{Duan:2005cp, EstebanPretel:2008ni}.

The discussion of the previous section remains unchanged except that
only the component of ${\bf B}$ along ${\bf P}$ contributes. This
amounts to using the effective oscillation frequency
$\gamma\cos2\theta$. One can verify in numerical examples that this
is indeed what happens. Even comparing our Figs.~\ref{fig:threeA}
and~\ref{fig:threeB} reveals to the naked eye that a large $\theta$
value increases the oscillation period.

We conclude that the essential dynamics of our system does not depend
on $\theta$. For our theoretical study it is most convenient to use
$\theta=0$ whereas in numerical examples one uses a nonvanishing
value that provides the necessary initial disturbance to start the
motion.

\subsection{Matter}

Ordinary matter has the same effect on all modes and therefore can be
removed by going into a rotating frame~\cite{Duan:2005cp}. Since we
anyway only study the internal motion of the ${\bf P}_i$ and since we
already go to a rotating frame to achieve this simplification,
nothing new happens by the extra matter-induced rotation. Therefore,
the effects discussed here are not modified by the presence of dense
matter except for details of the initial disturbance caused by the
fast-rotating ${\bf B}$.

\subsection{Antineutrinos}
\label{sec:antineutrinos}

Antineutrinos are most easily included in the EOMs as modes with
negative frequencies $\omega=-|\Delta m^2/2E|$ \cite{Raffelt:2007cb},
so in the most general case both neutrinos and antineutrinos are
present. On the other hand, we constantly switch between rotating
coordinate systems, shifting the zero-point of frequency at
convenience. After each shift the behavior of the abstract system is
the same, whereas its physical interpretation changes.

The polarization vectors used here are similar to the neutrino flavor
isospin vectors of Duan et~al.~\cite{Duan:2005cp} in that for an
antineutrino (negative frequency) ``spin up'' means $\bar\nu_\mu$ if
for a neutrino (positive frequency) ``spin up'' means $\nu_e$.
Shifting between rotating coordinate systems therefore changes, for
example, a $\bar\nu_\mu$ to~a~$\nu_e$.

We repeat, however, that the interpretation of the polarization
vectors is irrelevant with regard to the dynamics of the system. It is
easiest to think of all modes as representing neutrinos in that any
given distribution can be thought of as stemming from a neutrino
distribution, suitably shifted to a rotating frame. Therefore, in an
abstract discussion of the EOMs and their solutions, antineutrinos
need not appear explicitly.

\subsection{Fewer symmetries}

We have studied a very symmetric system, but the general behavior
persists if we vary the relative lengths of the vectors ${\bf
  P}_{\pm,0}$ and the frequency splittings. However, for any
motion of this sort to be possible the three vectors must be able to
move relative to each other while conserving ${\bf P}$. If ${\bf P}$
is conserved, energy conservation implies that we also need to
conserve~${\bf B}\cdot{\bf M}$.

Once the lengths of the three vectors are specified, their motion is
described by two angles each, so a total of six angles. Our
conditions provide four constraints, leaving two degrees of freedom:
an irrelevant overall precession angle around ${\bf P}$ and one angle
describing the internal configuration.

The general conditions for a solution are not particularly
illuminating, but we remark that, if ${\bf P}_+$ and ${\bf P}_-$ have
equal lengths, the parametric resonance requires $|{\bf
P}_0|<2\,|{\bf P}_\pm|$. This is verified in numerical examples.

\section{Spectrum of frequencies}                 \label{sec:spectrum}

\subsection{Numerical Example}

\begin{figure}[b]
\includegraphics[width=0.8\columnwidth]{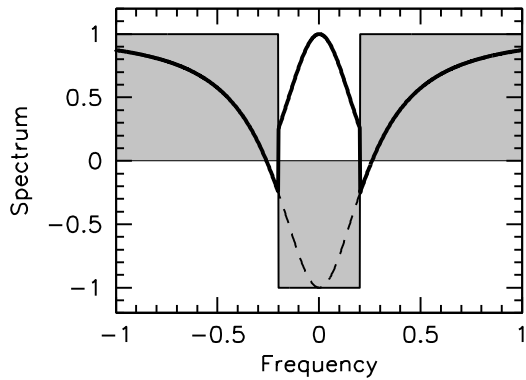}
\caption{Initial spectrum (thin line and shaded) and spectrum of
  maximum excursion (thick solid line), assuming a large fixed
  $\mu$. The dashed line is the ``flipped'' version of the thick
  spectrum.
\label{fig:resonance}}
\end{figure}

Next we consider a broad spectrum of modes. For convenience we scale
all frequencies with a scale $\gamma$ and use $\gamma t$ as a time
coordinate, i.e., we use dimensionless frequencies.  Without loss of
generality we consider the range $-1\leq\omega\leq+1$. We study the
continuous generalization of the previous three-vector example, using
an initial spectrum that is of box shape and has a flipped part in
the middle (shaded spectrum in Fig.~\ref{fig:resonance}). We let this
ensemble evolve with a fixed large~$\mu$, using a ${\bf B}$ that is
collinear with the polarization vectors except for a small mixing
angle that triggers the motion. All modes oscillate with the same
phase and reach their maximum excursion at the same time. We show the
maximum excursion spectrum as a thick line in
Fig.~\ref{fig:resonance}. The result is clearer when we ``flip'' the
central part of the spectrum (dashed line). Within numerical
accuracy, this compound curve is a Lorentzian resonance curve.

Therefore, the different parts of the neutrino spectrum librate
relative to each other. In particular, the neutrinos with
intermediate frequencies oscillate completely between two flavors
with a typical vacuum oscillation frequency. The neutrino density
does not enter as long as it is large in the sense $\mu\gg\gamma$.

We can make the ``flipped'' part of the spectrum narrower, obtaining
a narrower Lorentzian. However, we can not make it arbitrarily broad
in the same way as in the three-vector example the middle vector
could not be longer than twice the length of the peripheral vectors.
Here we find that the width of the flipped part must not exceed half
the total box width. The transition is sharp in the sense that if the
flipped part is slightly broader than half the total, nothing
happens, if it is slightly smaller we find complete reversals
and back of the central part of the spectrum.

\subsection{Analytic solution}

To determine the exact solution in the large-$\mu$ limit we note that
the EOMs in analogy to Eq.~(\ref{eq:EOM5}) are
\begin{equation}
\dot{\bf P}_\omega=\mu{\bf P}_\perp\times{\bf P}_\omega
+\omega{\bf B}\times{\bf P}_\omega\,,
\end{equation}
where each polarization vector is characterized by its frequency
$\omega$. We have assumed again that the mixing angle is small and
therefore ${\bf B}$ collinear with ${\bf P}=\int\D\omega\,{\bf
P}_\omega$. Integrating both sides over $\int\D\omega$ leads to
\begin{equation}
\dot{\bf P}_\perp={\bf B}\times({\bf M}-\mu\,P\,{\bf P}_\perp)
\end{equation}
where $P=|{\bf P}|$ is here not assumed to equal unity. The
``magnetic moment'' is ${\bf M}=\int\D\omega\,\omega\,{\bf
P}_\omega$. The same argument as in the three-vector example implies
${\bf P}_\perp={\bf M}/(\mu P)$ to lowest order in $\mu^{-1}$. We use
spectra that are symmetric relative to $\omega=0$ so that ${\bf M}$
to lowest order has no component along ${\bf B}$. So the EOMs are
\begin{equation}\label{eq:EOM10}
\dot{\bf P}_\omega=\left(\omega{\bf B}+\frac{{\bf M}}{P}
\right)\times{\bf P}_\omega\,.
\end{equation}
Once more we have eliminated the large frequency $\mu$.

In this form it is clear that we expect a resonance shape of the
oscillation pattern as a function of $\omega$. Only the mode with
$\omega=0$ is exactly on resonance in that the transverse $B$ field
(here ${\bf M}$) is exactly co-rotating with it, or in our
co-rotating coordinate system, does not move at all. The other modes
precess around ${\bf B}$ relative to the ${\bf M}$ direction. They
are not exactly on resonance.

For an explicit example we return to a box spectrum like
Fig.~\ref{fig:resonance} and define the function
\begin{equation}
s_\omega=\cases{+1&for $\beta<|\omega|\leq1$,\cr
-1&for $0\leq|\omega|\leq\beta$.\cr}
\end{equation}
The initial
polarization vectors are
\begin{equation}
{\bf P}_\omega=\pmatrix{0\cr0\cr s_\omega\cr}
\end{equation}
and so $P=|{\bf P}|=\int\D\omega\,s_\omega=2\,(1-2\beta)$. We further
define the Lorentzian function
\begin{equation}
f_\omega=\frac{1}{(\omega/\Gamma)^2+1}
\end{equation}
and fix its width by the condition
\begin{equation}\label{eq:zerocondition}
\int_{-1}^{+1}\D\omega\,f_\omega\,s_\omega=0
\end{equation}
which implies
\begin{equation}
\Gamma=\frac{\beta}{\sqrt{1-2\beta}}\,.
\end{equation}
This is only possible for $\beta<\frac{1}{2}$, precisely the case
where we find the parametric resonance.

As in the three-vector example we use a coordinate system where ${\bf
M}$ is oriented along the $y$ direction and the oscillation of ${\bf
P}_0$ takes place in the $x$-$z$ plane. Once more we describe this
vector by an angle $\varphi$ such that
\begin{equation}
{\bf P}_0=\pmatrix{\sin\varphi\cr0\cr-\cos\varphi\cr}\,.
\end{equation}
Inspired by the numerical example and by the symmetries of our system
we guess
\begin{equation}\label{eq:sol2}
{\bf P}_\omega=\left[\pmatrix{0\cr0\cr1\cr}
-\pmatrix{\sin\varphi\cr
(\omega/\Gamma)\,\sqrt{2(1-\cos\varphi)}
\cr1-\cos\varphi\cr}
f_\omega\right]\,s_\omega\,.
\end{equation}
Note that Eq.~(\ref{eq:zerocondition}) and the symmetry of $f_\omega$
and $s_\omega$ guarantee that ${\bf P}=\int\D\omega\,{\bf P}_\omega$
is the same for any~$\varphi$.

Next we integrate Eq.~(\ref{eq:EOM10}) over $\int\D\omega\,\omega$ so
that on the r.h.s.\ the term ${\bf M}\times{\bf M}$ drops out and we
find
\begin{equation}
\dot{\bf M}={\bf B}\times
\int\D\omega\,\omega^2\,{\bf P}_\omega\,.
\end{equation}
The vector ${\bf M}$ has only a $y$-component that is found to be
$\Gamma P\sqrt{2(1-\cos\varphi)}$. The vector on the r.h.s.\ also has
only a $y$-component $\Gamma^2P\sin\varphi$. Therefore, once more the
excursion angle evolves like an inverted pendulum,
\begin{equation}
\dot\varphi=\Gamma\,\sqrt{2(1-\cos\varphi)}\,.
\end{equation}
It is now easy to verify that Eq.~(\ref{eq:sol2}) indeed solves the
EOMs of Eq.~(\ref{eq:EOM10}).

\subsection{Single-crossed spectrum}

The behavior found in the previous section is similar for less
symmetric arrangements. However, the ``flipped'' part of the spectrum
must not be too broad and must not get too close to the edges of the
overall box. In other words, the two ``wings'' of the spectrum must
be large enough to support the parametric resonance, but we have not
worked out the general condition.

The crucial feature is to have a ``double-crossed'' spectrum
$s(\omega)$, representing the initial $z$-component of ${\bf
P}_\omega$. The spectrum must first cross from positive to negative
and then back to positive values or the other way round. One can also
construct multiple-crossed spectra and finds more complicated
parametric resonance patterns.

On the other hand, this form of collective motion can not occur for a
single-crossed spectrum where $s(\omega)<0$ for $\omega<\omega_{\rm
cross}$ and $s(\omega)>0$ for $\omega>\omega_{\rm cross}$. Here ${\bf
B}\cdot{\bf M}$ is maximal, so moving any polarization vector makes
it smaller. This is easily seen if we go to a rotating frame where
$\omega_{\rm cross}=0$. This has the effect ${\bf B}\cdot{\bf M}={\bf
B}\cdot\sum_i\omega_i{\bf P}_i\to {\bf
B}\cdot\sum_i(\omega_i-\omega_{\rm cross}){\bf P}_i={\bf B}\cdot{\bf
M}-\omega_{\rm cross}{\bf B}\cdot{\bf P}$ and since ${\bf B}\cdot{\bf
P}$ is a constant of the motion, we have only added a constant term
to the energy. In the new system with $\omega_{\rm cross}=0$ all
modes with negative frequencies get multiplied with a negative
$s(\omega)$, the positive ones with a positive $s(\omega)$.
Therefore, any motion of ${\bf P}_\omega$ lowers ${\bf B}\cdot{\bf
M}$. The opposite is true if the spectrum crosses from positive to
negative values or if we take ${\bf B}$ to point in the negative
$z$-direction.  In other words, even though for a single-crossed
spectrum one could move the polarization vectors such that $|{\bf
P}|$ and $(\mu/2)\,{\bf P}^2$ remain unchanged, the quantity ${\bf
B}\cdot{\bf M}$ is extremal.

A single-crossed spectrum is the prototype for a gyroscopic flavor
pendulum that is driven by the exchange of energy between
$(\mu/2)\,{\bf P}^2$ (playing the role of kinetic energy) and ${\bf
B}\cdot{\bf M}$ (potential energy). In the large-$\mu$ limit the
oscillations are synchronized and, if the mixing angle is small,
nothing visible happens.

If one assumes that a source only emits neutrinos and antineutrinos
of one flavor, e.g.\ a flux of $\nu_e$ (positive $\omega$) together
with a flux of $\bar\nu_e$ (negative $\omega$) one has a
single-crossed spectrum in that the polarization vectors for $\nu_e$
are ``spin~up,'' those for $\bar\nu_e$ ``spin~down'' as explained in
Sec.~\ref{sec:antineutrinos}. Therefore, in this situation the
parametric resonance does not play any role.

\subsection{Nonisotropic system}
\label{sec:multiangle}

The parametric resonance is probably not possible if the neutrino
ensemble is not isotropic. Assuming axial symmetry around some
direction we can describe every mode by its frequency $\omega$ and
the velocity projection $v$ on the symmetry axis. The EOMs are in
this case~\cite{Duan:2006an, Raffelt:2007yz}
\begin{equation}\label{eq:EOM21}
 \dot {\bf P}_{\omega,v}=
 {\bf H}_{\omega,v}\times {\bf P}_{\omega,v}\,,
\end{equation}
where
\begin{equation}\label{eq:ham21}
{\bf H}_{\omega,v}=\omega{\bf B}
+\mu({\bf P}-v{\bf F})
\end{equation}
and
\begin{equation}\label{eq:ham22}
{\bf P}=\int\D\omega\,\D v\,{\bf P}_{\omega,v}\,,
\quad\quad
{\bf F}=\int\D\omega\,\D v\,v\,{\bf P}_{\omega,v}\,.
\end{equation}
Even if the vectors ${\bf P}$ and ${\bf F}$ are collinear, there is
no longer a co-rotating frame where all polarization vectors would
see essentially the same effective magnetic field. If one of them
were on resonance with regard to some transverse field, the others
would precess with different angular velocities where the differences
are much larger than order $\omega$ in the large-$\mu$ limit.

We have studied a few numerical examples where we assumed a
homogeneous ensemble with a ``half-isotropic'' distribution, i.e.,
all modes with velocity components in the positive direction of the
symmetry axis were isotropically occupied, those with negative
velocity components were taken to be empty. The parametric resonance
indeed disappeared for all tested examples where it occurred in the
corresponding isotropic case.

\section{Discussion}                          \label{sec:conclusions}

We have shown that in a dense neutrino gas, neutrino-neutrino
interactions can lead to a parametric resonance that causes different
parts of the flavor spectrum to oscillate relative to each other.
This effect persists in the large-density limit where one would have
expected synchronized oscillations. The new pattern is an internal
motion among the ensemble of polarization vectors that still precess
together as a collective object, but do not retain a common
orientation relative to each other. The strength of the
neutrino-neutrino interaction does not appear in the oscillation
frequency that is determined entirely by the spectrum of vacuum
oscillation frequencies.

However, the parametric resonance has a macroscopic impact only if
the initial flavor spectrum $s(\omega)$ is at least ``double
crossed'' in that there must be a spectral flavor sequence of the
form up-down-up or the other way round. We use polarization vectors
similar to ``neutrino flavor isospin vectors'' where, for example, a
$\bar\nu_e$ mode is ``spin down'' if $\nu_e$ is defined as
``spin~up.'' Therefore, a neutrino flux initially consisting of
neutrinos and antineutrinos of a single flavor represents a
single-crossed case: $s(\omega)$ changes sign once at $\omega=0$
going from antineutrinos to neutrinos. Most numerical studies of
supernova neutrino oscillations used such single-crossed spectra and
the parametric resonance did not show up.

A dense gas of neutrinos and antineutrinos of different flavors in
kinetic (but not chemical) equilibrium is also single crossed. For a
Fermi-Dirac distribution the spectral shape and amplitude are fixed
by the temperature and chemical potential. If the distributions for
two flavors are described by the same $T$ but different degeneracy
parameters, they do not cross. Our spectrum $s(\omega)$ with
$\omega=\Delta m^2/2E$ represents the difference between the
distributions of the two flavors because the identical parts drop out
of the oscillation equations. The only crossing occurs at $\omega=0$
at the spectral junction between antineutrinos and neutrinos.

Flavor-dependent neutrino chemical potentials are important inside a
supernova core and also in the early universe if primordial neutrino
asymmetries exist. The usual picture of synchronized oscillations
and/or a synchronized MSW resonance remains unchanged. In this
context the parametric resonance apparently does not lead to a novel
speed--up of flavor conversions.

Neutrinos streaming from a supernova core are not black-body
radiation. Even if one approximates the fluxes of $\nu_e$,
$\bar\nu_e$ and the other species by Fermi-Dirac distributions, the
effective temperatures and chemical potentials are different.
Typically one finds three spectral crossings, the usual one at
$\omega=0$ and one for a negative and one for a positive $\omega$,
see for example Ref.~\cite{Fogli:2007bk}. However, neutrinos
streaming from a source are strongly anisotropic, so the parametric
resonance would be suppressed by multi-angle effects as discussed in
Sec.~\ref{sec:multiangle} even if the spectral conditions were
appropriate.

Therefore, it is unclear if the collective motion represented by the
parametric resonance is realized in any practical astrophysical or
cosmological context. Perhaps the main insight is that even for a
slowly changing or fixed $\mu$ the individual polarization vectors
${\bf P}_i$ need not follow their ``single-particle Hamiltonians''
${\bf H}_i$ even if the system as a whole evolves adiabatically.

In the supernova context one is primarily concerned with the fate of
flavor-dependent neutrino spectra as a function of radius, the
decreasing neutrino-neutrino interaction producing a ``spectral
split.'' In the homogeneous and isotropic model represented by
Eq.~(\ref{eq:EOM1}) this occurs if $\mu$ slowly decreases so that
${\bf H}_i$ changes from being $\mu{\bf P}$--dominated to being
$\omega_i{\bf B}$--dominated. For a single-crossed spectrum the
occurrence of a split is explained by each ${\bf P}_i$ following its
${\bf H}_i$ in the co-rotating frame. The spectral split occurs at a
frequency $\omega_{\rm split}$ corresponding to the final
co-rotation frequency. Its value can be found from the conservation
of ${\bf B}\cdot{\bf P}$. In the present case of a double-crossed
spectrum a slowly decreasing $\mu$ leads to two spectral splits.
Multiple-crossed spectra lead to multiple spectral splits. The
occurrence of multiple splits is not directly accounted for by the
picture of all ${\bf P}_i$ following their ${\bf H}_i$ in a single
co-rotating frame and a full theoretical understanding of the
phenomenon of multiple spectral splits is presently missing.

The parametric resonance and multiple splits are not directly
related even though this investigation was motivated by the
numerical observation of multiple splits. A parametric resonance for
our box-spectrum requires the flipped middle part not to exceed half
the box width whereas a double split occurs for any width of the
flipped part. It is actually surprising that the large-$\mu$
oscillations have no apparent impact on the final sharp
double--split that is found in the adiabatic limit of a slowly
decreasing $\mu$. One also finds that multiple splits are not, or at
least not always, prevented by multi-angle effects for neutrinos
streaming from a source. On the other hand, the collective motion
found here is ``fragile'' and easily suppressed by multi-angle
effects as explained earlier.

Perhaps the main lesson from our investigation is that the
innocent-looking EOMs of Eq.~(\ref{eq:EOM1}) continue to surprise us
with unexpected solutions, even if these solutions need not be of
direct astrophysical relevance. Much of the insights gained about
collective neutrino oscillations are owed to ``numerical
experiments'' followed in some cases by theoretical explanations and
even analytic solutions. One may well worry if our current
understanding of these nonlinear equations is sufficiently advanced
to arrive at robust conclusions about possible observable effects
for example in supernova neutrino spectra. More theoretical work may
well turn up more surprises.

\section*{Acknowledgements} 

This work was partly supported by the Deutsche Forschungsgemeinschaft
under grant TR-27 ``Neutrinos and Beyond'' and the Cluster of
Excellence ``Origin and Structure of the Universe'' (Munich and
Garching).


\end{document}